\documentclass[12pt]{article}
\usepackage{graphics}
\usepackage{epsf}
\begin{document}
\begin{titlepage}
\def\baselinestretch{1.2}
\topmargin     -0.25in

\rightline{LAPTH-636/98}
 \rightline{hep-ph/9809220}
\rightline{May 1998}

\vspace*{2cm}
\begin{center}
{\large {\bf Physics at the Linear Collider}}\footnote{Invited talk given at the Fifth
Workshop on High Energy Physics Phenomenology, Inter-University Centre for Astronomy and
Astrophysics, Pune, India, January 12 - 26, 1998.}

\vspace*{0.5cm}

{\bf Fawzi Boudjema }
\\

{\it  Laboratoire de Physique Th\'eorique} {\large
LAPTH}\footnote{URA 14-36 du CNRS, associ\'ee \`a l'Universit\'e
de Savoie.}\\ {\it Chemin de Bellevue, B.P. 110, F-74941
Annecy-le-Vieux, Cedex, France.}

\end{center}

\centerline{ {\bf Abstract} }
\baselineskip=14pt
\noindent
 {\small The physics at the planned  $e^+e^-$ colliders is discussed around three main topics
 corresponding to different manifestations of symmetry breaking: $W$ physics in the no Higgs scenario,
 studies of the properties of the Higgs and precision tests of SUSY. A comparison with the LHC is made
 for all these cases. The $\gamma \gamma$ mode of the linear collider will also be
 reviewed.}

\vspace{1.5cm}

\hrule {\bf Keywords:} {\small {\it Linear Colliders,
Photon-Photon Colliders, Polarization, Symmetry Breaking, Chiral
Lagrangians, Supersymmetry, Higgs Bosons.}}




\end{titlepage}
\baselineskip=18pt


\newcommand{\be}{\begin{equation}}
\newcommand{\beq}{\begin{equation}}
\newcommand{\eeq}{\end{equation}}
\newcommand{\ee}{\end{equation}}

\newcommand{\beqn}{\begin{eqnarray}}
\newcommand{\eeqn}{\end{eqnarray}}
\newcommand{\bea}{\begin{eqnarray}}
\newcommand{\ena}{\end{eqnarray}}
\newcommand{\ra}{\rightarrow}
\newcommand{\susy}{{\cal SUSY$\;$}}
\newcommand{\su}{$ SU(2) \times U(1)\,$}

\newcommand{\gag}{$\gamma \gamma$ }
\newcommand{\gam}{\gamma \gamma }
\def\W{{\mbox{\boldmath $W$}}}
\def\B{{\mbox{\boldmath $B$}}}
\newcommand{\np}{Nucl.\,Phys.\,}
\newcommand{\pl}{Phys.\,Lett.\,}
\newcommand{\pr}{Phys.\,Rev.\,}
\newcommand{\prl}{Phys.\,Rev.\,Lett.\,}
\newcommand{\prep}{Phys.\,Rep.\,}
\newcommand{\zp}{Z.\,Phys.\,}
\newcommand{\sovjnp}{{\em Sov.\ J.\ Nucl.\ Phys.\ }}
\newcommand{\nuclinst}{{\em Nucl.\ Instrum.\ Meth.\ }}
\newcommand{\annp}{{\em Ann.\ Phys.\ }}
\newcommand{\intjmp}{{\em Int.\ J.\ of Mod.\  Phys.\ }}

\newcommand{\eps}{\epsilon}
\newcommand{\mw}{M_{W}}
\newcommand{\mww}{M_{W}^{2}}
\newcommand{\mwmw}{M_{W}^{2}}
\newcommand{\mhmh}{M_{H}^2}
\newcommand{\mz}{M_{Z}}
\newcommand{\mzz}{M_{Z}^{2}}

\newcommand{\lra}{\leftrightarrow}
\newcommand{\tr}{{\rm Tr}}
\def\ls1{{\not l}_1}
\newcommand{\cms}{centre-of-mass\hspace*{.1cm}}

\newcommand{\dkg}{\Delta \kappa_{\gamma}}
\newcommand{\dkz}{\Delta \kappa_{Z}}
\newcommand{\dz}{\delta_{Z}}
\newcommand{\dgz}{\Delta g^{1}_{Z}}
\newcommand{\dgzt}{$\Delta g^{1}_{Z}\;$}
\newcommand{\la}{\lambda}
\newcommand{\lag}{\lambda_{\gamma}}
\newcommand{\lambdae}{\lambda_{e}}
\newcommand{\laz}{\lambda_{Z}}
\newcommand{\lnl}{L_{9L}}
\newcommand{\lnr}{L_{9R}}
\newcommand{\lt}{L_{10}}
\newcommand{\lu}{L_{1}}
\newcommand{\ld}{L_{2}}
\newcommand{\cw}{\cos\theta_W}
\newcommand{\sw}{\sin\theta_W}
\newcommand{\tw}{\tan\theta_W}

\newcommand{\epm}{$e^{+} e^{-}\;$}
\newcommand{\epemt}{$e^{+} e^{-}\;$}
\newcommand{\epem}{e^{+} e^{-}\;}
\newcommand{\ememt}{$e^{-} e^{-}\;$}
\newcommand{\emem}{e^{-} e^{-}\;}
\newcommand{\eeww}{e^{+} e^{-} \ra W^+ W^- \;}
\newcommand{\eewwt}{$e^{+} e^{-} \ra W^+ W^- \;$}
\newcommand{\epemww}{e^{+} e^{-} \ra W^+ W^- }
\newcommand{\epemwwt}{$e^{+} e^{-} \ra W^+ W^- \;$}
\newcommand{\eennhht}{$e^{+} e^{-} \ra \nu_e \bar \nu_e HH\;$}
\newcommand{\eennhh}{e^{+} e^{-} \ra \nu_e \bar \nu_e HH\;}
\newcommand{\ppwg}{p p \ra W \gamma}
\newcommand{\wwhh}{W^+ W^- \ra HH\;}
\newcommand{\wwhht}{$W^+ W^- \ra HH\;$}
\newcommand{\ppwz}{pp \ra W Z}
\newcommand{\ppwgt}{$p p \ra W \gamma \;$}
\newcommand{\ppwzt}{$pp \ra W Z \;$}
\newcommand{\gamgamt}{$\gamma \gamma \;$}
\newcommand{\gamgam}{\gamma \gamma \;}
\newcommand{\egamt}{$e \gamma \;$}
\newcommand{\egam}{e \gamma \;}
\newcommand{\gamgamwwt}{$\gamma \gamma \ra W^+ W^- \;$}
\newcommand{\gamgamwwht}{$\gamma \gamma \ra W^+ W^- H \;$}
\newcommand{\gamgamwwh}{\gamma \gamma \ra W^+ W^- H \;}
\newcommand{\gamgamwwhht}{$\gamma \gamma \ra W^+ W^- H H\;$}
\newcommand{\gamgamwwhh}{\gamma \gamma \ra W^+ W^- H H\;}
\newcommand{\ggww}{\gamma \gamma \ra W^+ W^-}
\newcommand{\ggwwt}{$\gamma \gamma \ra W^+ W^- \;$}
\newcommand{\ggwwht}{$\gamma \gamma \ra W^+ W^- H \;$}
\newcommand{\ggwwh}{\gamma \gamma \ra W^+ W^- H \;}
\newcommand{\ggwwhht}{$\gamma \gamma \ra W^+ W^- H H\;$}
\newcommand{\ggwwhh}{\gamma \gamma \ra W^+ W^- H H\;}
\newcommand{\ggwwz}{\gamma \gamma \ra W^+ W^- Z\;}
\newcommand{\ggwwzt}{$\gamma \gamma \ra W^+ W^- Z\;$}
\def\smx{{\cal{S}} {\cal{M}}\;}

\newcommand{\ptu}{p_{1\bot}}
\newcommand{\vecptu}{\vec{p}_{1\bot}}
\newcommand{\ptd}{p_{2\bot}}
\newcommand{\vecptd}{\vec{p}_{2\bot}}
\newcommand{\ie}{{\em i.e.}}
\newcommand{\cm}{{{\cal M}}}
\newcommand{\cl}{{{\cal L}}}
\newcommand{\cd}{{{\cal D}}}
\newcommand{\cv}{{{\cal V}}}
\def\slashc{c\kern -.400em {/}}
\def\slashL{L\kern -.450em {/}}
\def\slashcl{\cl\kern -.600em {/}}
\def\Ww{{\mbox{\boldmath $W$}}}
\def\B{{\mbox{\boldmath $B$}}}
\def\noi{\noindent}
\def\nn{\noindent}
\def\sm{${\cal{S}} {\cal{M}}\;$}
\def\nph{${\cal{N}} {\cal{P}}\;$}
\def\sb{$ {\cal{S}}  {\cal{B}}\;$}
\def\ssb{${\cal{S}} {\cal{S}}  {\cal{B}}\;$}
\def\ssbe{{\cal{S}} {\cal{S}}  {\cal{B}}}
\def\cviol{${\cal{C}}\;$}
\def\pviol{${\cal{P}}\;$}
\def\cpviol{${\cal{C}} {\cal{P}}\;$}

\newcommand{\lgg}{\lambda_1\lambda_2}
\newcommand{\lww}{\lambda_3\lambda_4}
\newcommand{\ppin}{ P^+_{12}}
\newcommand{\pmin}{ P^-_{12}}
\newcommand{\ppout}{ P^+_{34}}
\newcommand{\pmout}{ P^-_{34}}
\newcommand{\sinsq}{\sin^2\theta}
\newcommand{\cossq}{\cos^2\theta}
\newcommand{\yt}{y_\theta}
\newcommand{\hppll}{++;00}
\newcommand{\hpmll}{+-;00}
\newcommand{\hpplt}{++;\lambda_30}
\newcommand{\hpmlt}{+-;\lambda_30}
\newcommand{\hpptt}{++;\lambda_3\lambda_4}
\newcommand{\hpmtt}{+-;\lambda_3\lambda_4}
\newcommand{\dk}{\Delta\kappa}
\newcommand{\klam}{\Delta\kappa \lambda_\gamma }
\newcommand{\kac}{\Delta\kappa^2 }
\newcommand{\lac}{\lambda_\gamma^2 }
\def\gamgamtzz{$\gamma \gamma \ra ZZ \;$}
\def\gamgamtww{$\gamma \gamma \ra W^+ W^-\;$}
\def\gamgamtwwe{\gamma \gamma \ra W^+ W^-}

\setcounter{section}{1}

\setcounter{subsection}{0}
\setcounter{equation}{0}
\def\thesubsection {\thesection.\arabic{subsection}}
\def\theequation{\thesection.\arabic{equation}}

\setcounter{equation}{0}
\def\thequation{\thesection.\arabic{equation}}

\setcounter{section}{0}
\setcounter{subsection}{0}

\section{Introduction}
\subsection{The projects and the designs}
 There has been an intense activity during the last
decade in the physics of a high energy \epemt linear collider. Several working groups in
Europe, the USA and Japan have been set up. These groups have on the one hand addressed
and tackled the feasibility and construction of such a machine, and on the other have by
now convincingly made a strong point as concerns the advantages and benefits that such a
collider can bring to our understanding of the fundamental issue in Physics: the
mechanism of symmetry breaking (\ssb) and the concomitant mass problem. In Europe, for
instance, since 1991 five one-year-long Workshops have been organized, with three general
meetings each\cite{DesyNLC}. The end of each of these Workshops has coincided with an
international \epemt linear collider meeting where the studies of various groups in
Japan, the US and Europe are summarized, compared and
complemented\cite{eeInternationalWorkshop}. Along side, the machine people who have been
working on different designs have had regular international meetings.

There is general consensus for a machine which in a first stage
would run around $500$GeV or at the top threshold with a
luminosity of $10-80fb^{-1}$, and which should be upgraded to
1-2TeV. This means that ideally one should, from the start, have a
machine with a length (~15-30kms) so that there is enough space
for the later adjunction of more accelerating devices which allow
to reach the TeV regime. At the same time one has to increase the
luminosity as the energy increases, to make up for the falling
cross sections. Also, although the reason for building a linear
rather than a circular collider \epemt is to avoid the prohibitive
synchroton energy loss, there is nonetheless some energy loss due
to beamstrahlung. This is a coherent radiation which occurs as a
result of each beam feeling the intense electromagnetic field
created by the opposite tightly dense bunch. If this radiation is
not controlled, the huge photon flux (and accompanying \epemt
pairs) will create a dirty background much like in a hadron
collider. Energy, luminosity and beamstrahlung are the key
parameters that enter in the designs of the various proposals and
set constraints on their parameters(see for example the technical
design reports\cite{DESY-TESLA,NLCZDR} and
also\cite{eePeskinReview}). Striving to have a luminosity,
${\cal{L}}$ increasing as $s$ (the cms energy squared), one should
arrange to have beams with very small spot-sizes ${\sigma_{x,y}}$.
However, this situation also leads to large beamstrahlung. One
then has to find a compromise and allow for instance for flat
beams ${\sigma_x} \gg {\sigma_y}$. This compromise is reflected in
the formula for the luminosity:

\beq
 {\cal{L}} \propto (n N f E) \times \frac{N}{\sigma_x} \times
\frac{1}{\sigma_y} \propto {\cal{P}} \times n_\gamma \times
\frac{1}{\sigma_y} \eeq

\noi $N$ is the number of particles per bunch, $n$ the number of
bunches,$f$ the RF frequency. The first factor is the beam power
${\cal{P}}$, the second gives $n_\gamma$ the number of
beamstrahlung photons that should be kept to a minimum.

Another important feature of the linear collider is the
availability of polarization. $95\%$ degree of polarization for
the electron beam is foreseen (note that SLD at SLAC has already
achieved about $80\%$ beam polarization). Some studies have also
shown that one could polarize the positrons ($60\%$ seems
possible).

\begin{figure*}[hbtp]
\caption{\label{Clic1TeV}{\em The 1TeV Clic Complex (from the CLIC
homepage\cite{homepagesLC}).}}
\begin{center}
 \mbox{\epsfxsize=16.5cm\epsfysize=11.5cm\epsffile{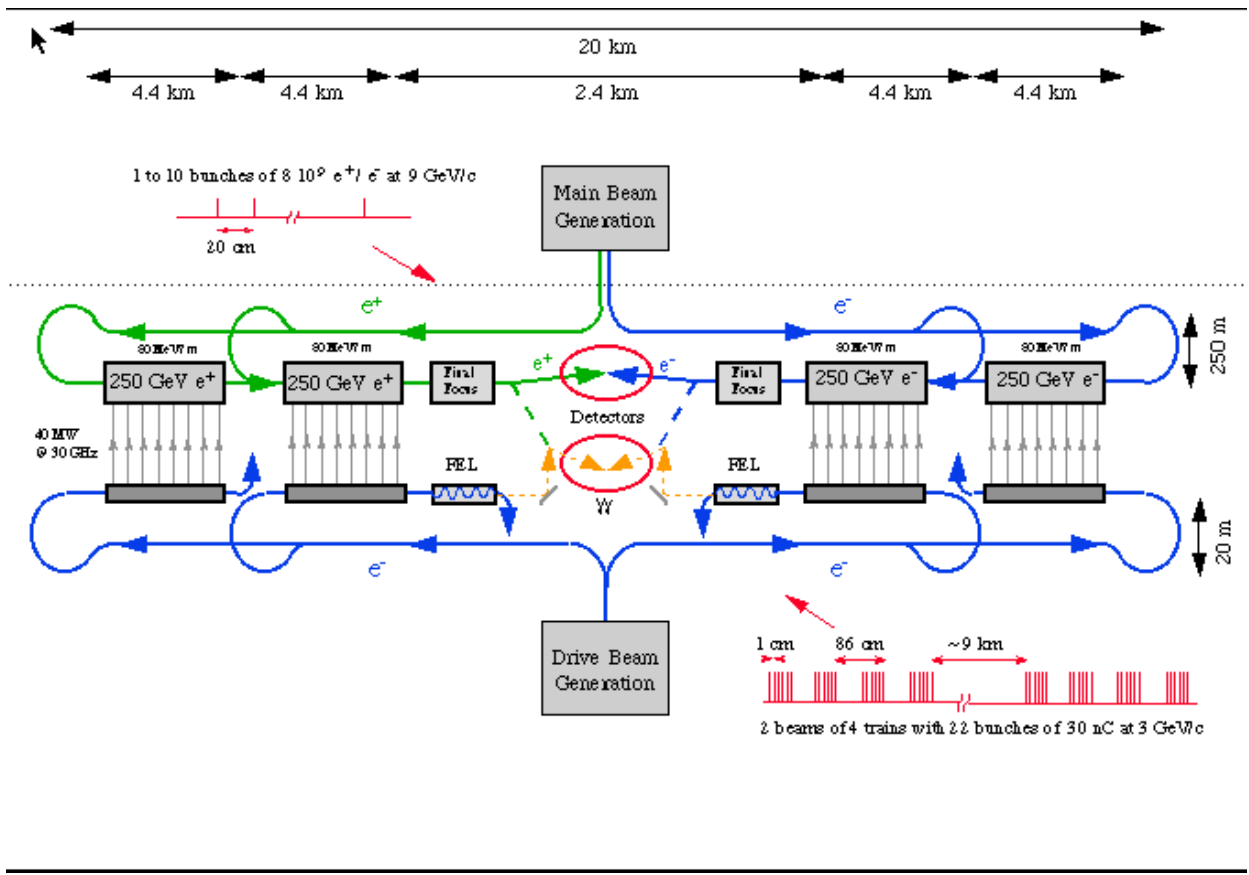}}
\end{center}
\end{figure*}

The main designs (refer to the corresponding
homepages\cite{homepagesLC}) have been developed at DESY (Tesla
which relies on a superconducting structure and the S-band SBLC),
KEK (JLC with an option of running up to 1.5-2TeV), SLAC (NLC).
All of these projects have had some test facilities and have also
joined effort like with the international collaboration which
successfully tested the final focus (FFTB: final focus tests
facility). CERN has also a very ambitious project (CLIC), while
with the financial crisis that has terribly hit Russian research,
the Protvino project (VLEPP) will most probably never be realized.
A layout of the CLIC design is shown in Fig.~\ref{Clic1TeV}.

\subsection{A \gamgamt and \egamt Collider}
\begin{figure*}[hbtp]
\caption{\label{lasernew}{\em The laser scheme of converting an
electron of the linac into a highly energetic photon (see text).}}
\begin{center}
\vspace*{-1.5cm} \mbox{\epsfxsize=6.5cm\epsfysize=6.5cm\epsffile{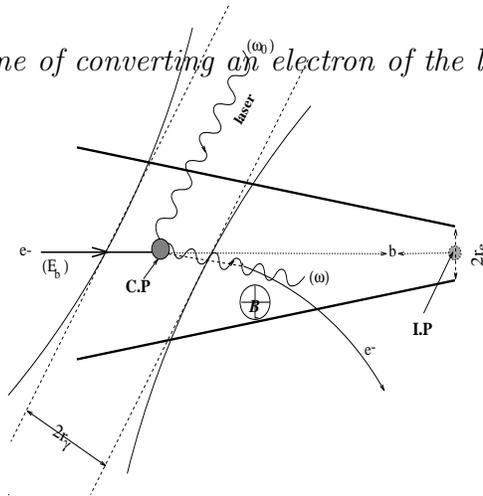}}
\end{center}
\end{figure*}

As can be seen from Fig.~\ref{Clic1TeV}, the CLIC design allows for a second interaction
region devoted to \gamgamt collisions. This is now an option which is taken seriously, as
an add-on, in all designs. Since the organizers have asked me to spend some time on this
option and since some of the Working Groups will look into the physics at these new
colliders, I shall comply by going through some detail. \noi Apart from the almost
straightforward possibility to turn the \epemt machine into an \ememt mode, there is the
very exciting prospect to convert either one beam of the machine or both into an intense
and collimated photon beam thus turning the machine into a \egamt or \gamgamt
collider\cite{PhotonCol,ggTESLA}. The idea, see Fig.~\ref{lasernew}, is to focus an
intense laser beam (with a frequency corresponding to a few eV) at an extremely small
angle onto the single pass electron. At some conversion point (CP) a few centimeters away
from the interaction point (IP) the laser photon Compton backscatters on the single-pass
electron with the result that most of the energy of the electron, $E_b$, gets transferred
to the photon beam. The latter then reaches the IP with a spread of the order of that of
the original electron.
The remaining soft electron from the conversion can nonetheless be
an nuisance. Not only it will scatter a few times with the photon
and therefore distorts the spectrum but it can also make it to the
IP. In this case the initial state would be a mixture of \epemt,
\egamt and \gamgamt thus creating again an unwanted background.
One suggestion\cite{PhotonCol} is to simply sweep these remaining
soft electrons by applying a strong transversal magnetic field
(~1T) within the space between the IP and the CP. But then it is
still not clear what (damaging) effect this will have on the
detectors especially the microvertex detector which is so crucial
for Higgs studies. \\ \noi A key parameter of the \gamgamt
collider, $x_0$, is directly related to the maximum energy,
$\omega_{max}$ that can be taken up by the photon. It is
introduced through the scaled invariant mass of the original
$e\gamma$ system and for a head-on hit of the laser is given by:

\beq x_0=\frac{M_{e\gamma_0}^2}{m_e^2}-1=\frac{4 E_b
\omega_0}{m_e^2} \simeq (15.3) \left((\frac{E_b}{TeV})\right)
\left(\frac{\omega_0}{eV}\right) \;\;\; \mbox{{\rm so that }}\;\;
\omega_{\max.}=\frac{x_0}{x_0+1} E_b \label{omemax} \eeq

Most of the photons are emitted at extremely small angles with the
most energetic photons scattered at zero angle. With the typical
angle $\theta_0=(m_e/E_b) \sqrt{x+1}$ of order some $\mu$rd, the
spread of the high-energy photon beam is thus of order some 10's
{\em nm}. The energy spread is roughly given by $\omega\approx
\omega_{max}/ (1+(\theta/\theta_0)^2)$.
 It is clear that the further away from the I.P. the conversion occurs,
those photons that make it to the I.P. are those with the smallest
scattering angle and hence with the maximum energy. These are the
ones that will contribute most to the luminosity. Therefore, with
a large distance of conversion one has a high monochromaticity at
the expense of a small integrated (over the energy spectrum)
luminosity. For those processes whose cross-section is largest for
the highest possible energy, this particular set-up would be
advantageous especially in reducing possible backgrounds that
dominate at smaller invariant \gag masses.

\noi From Eq.~\ref{omemax} it is clear that in order to reach the
highest possible photon energies one should aim at having as large
a $x_0$ as possible. However, one should be careful that the
produced photon and the laser photon do not interact so that they
create a \epm pair (first threshold); the laser frequency should
be chosen or tuned such that one is below the \epm threshold. If
we want maximum energy, it is by far best to choose the largest
$x_0$ taking into account this restriction. The optimal $x_0$ is
then given by $x_0 \leq 2 (1+\sqrt{2}) \sim 4.83$. This value
means that the photon can take up as much as $83\%$ of the beam
energy. \noi Naturally, the luminosity spectrum depends directly
on the differential Compton cross-section. The original electron
as well as the laser can be polarized\cite{PhotonCol}, resulting
in quite distinctive spectra depending on how one chooses the
polarizations. \noi The \gag luminosity spectrum is a convolution
involving the differential Compton cross-sections of the two
photons
 as well as a
conversion function that depends very sensitively on the
conversion distances and the characteristics of the linac beams.
The energy dependence of the former function is only through the
energy fraction $\sqrt{\tau}$, while the conversion function
involves the \epm cm energy explicitly. Realistically other
considerations should be taken into account. These have to do with
the laser power. In most theoretical studies it has been assumed
that the density of the laser photons is such that all the
electrons are converted (this assumes a conversion coefficient,
$k=1$) and that multiple scattering is negligible. A compact
analytical form for the conversion function is obtained in the
case of a Gaussian profile for the electron beam with an azimuthal
symmetry. Moreover almost all the physics analyses have been done
with $b=0$. Before tackling more realistic spectra, it is worth
reviewing the properties of these spectra in the simple case (with
analytical formulae), in order to exhibit the importance of
polarisation.
\begin{figure*}[hbtp]
\caption{\label{spectre12}{\bf (a)} {\em The total luminosity
spectra in the case of different combinations of the longitudinal
polarizations of the linac electrons and the circular
polarizations of the laser. The ``classic" Weisz\"acker-Williams
spectrum is shown for comparison. The spectra assume a distance of
conversion, $b=0$.} {\bf (b)} {\em Projecting the contributions of
the $J_Z=0$ and the $J_Z=2$ polarized spectrum in the peaked
spectrum setting $2\lambda_e P_c=2 \lambda_e' P_c'=-1$.}}
\begin{center}
\vspace*{-1.5cm}
\mbox{\epsfxsize=14cm\epsfysize=9cm\epsffile{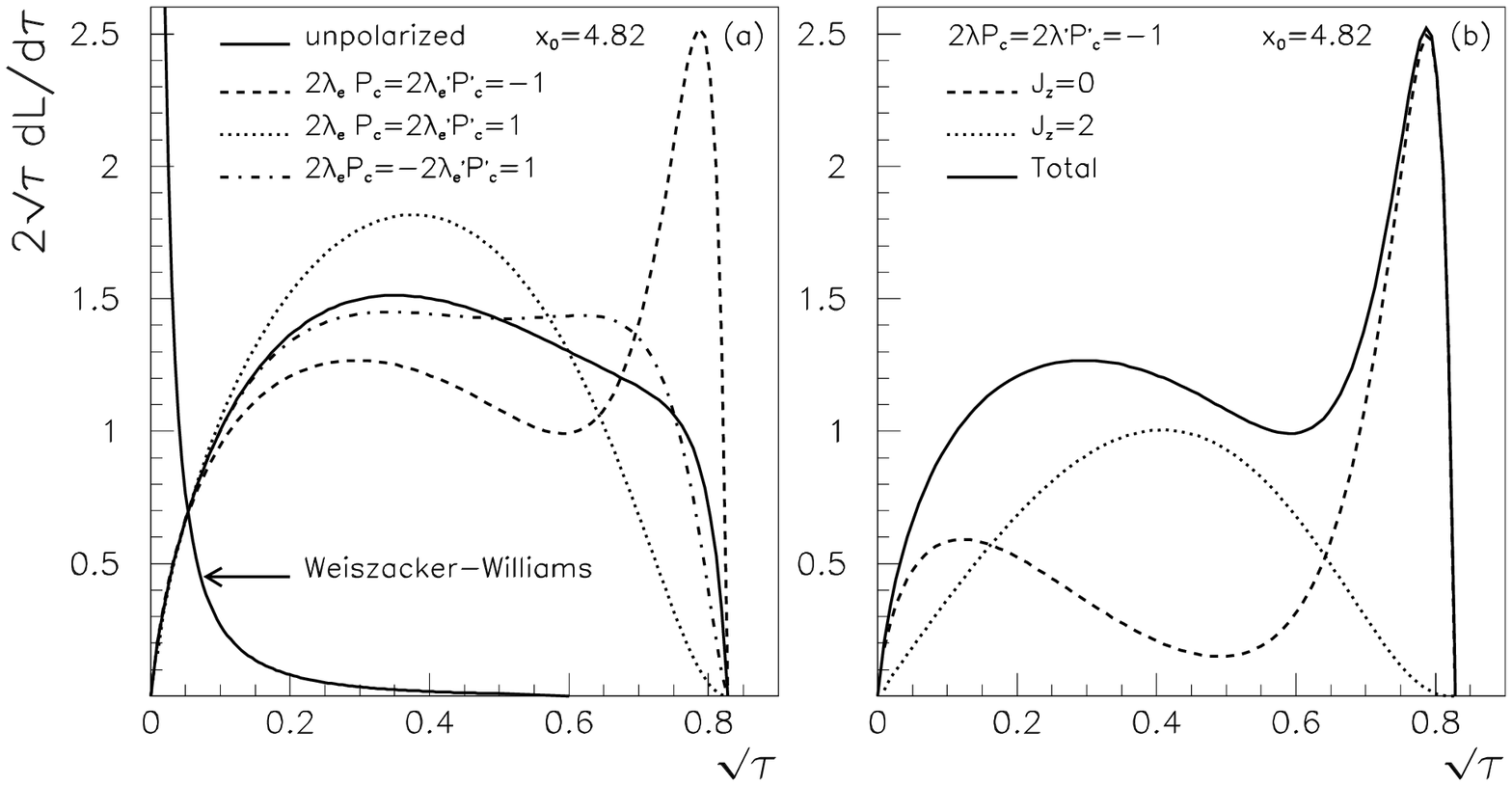}}
\vspace*{-1.cm}
\end{center}
\end{figure*}

\noi In Fig.~\ref{spectre12}a we compare the luminosity spectrum
(as a function of the reduced \gag invariant mass) that one
obtains by choosing different sets of polarizations for the two
arms of the photon collider. First of all, in all cases and as
advertised earlier one has a hard spectrum compared to the
``classic" Weisz\"acker-Williams spectrum. In case of no
polarization at all, one obtains a broad spectrum which is almost
a step function that extends nearly all the way to the maximum
energy (restricted by the value of $x_0$). The hardest spectrum is
arrived at by choosing the circular polarization of the laser
($P_c$) and the mean helicity of the electron ($\lambda_e$) to be
opposite, i.e., $2 \lambda_e P_c=-1$, for {\it both} arms of the
collider. In the case where both arms have $2 \lambdae P_c=+1$ the
spectrum has a ``bell-like" shape which favours the middle range
values of $\sqrt\tau$. In the case where the two arms of the
collider have an opposite value for the product $2\lambdae P_c$,
the spectrum is almost identical to the one obtained in case of no
polarization.
\begin{figure*}[hbtp]
\caption{\label{spectre34}{\em {\bf (a)} Projecting the contributions of the $J_Z=0$ and
the $J_Z=2$ polarized spectrum in the ``broad" setting $2\lambda_e P_c=2\lambda_e'
P_c'=1$ (with a conversion distance $b=0$). Thick lines are with a $100\%$ longitudinal
polarization for the electron while the light lines are for $50\%$ longitudinal
polarization. The lasers are taken to be fully right-handed. {\bf (b)} As in {\bf (a)}
but for unpolarized electrons and where we have imposed a rapidity cut of $\eta<1$.}}
\vspace{-1.5cm}
\begin{center}
 \mbox{\epsfxsize=14cm\epsfysize=9.cm\epsffile{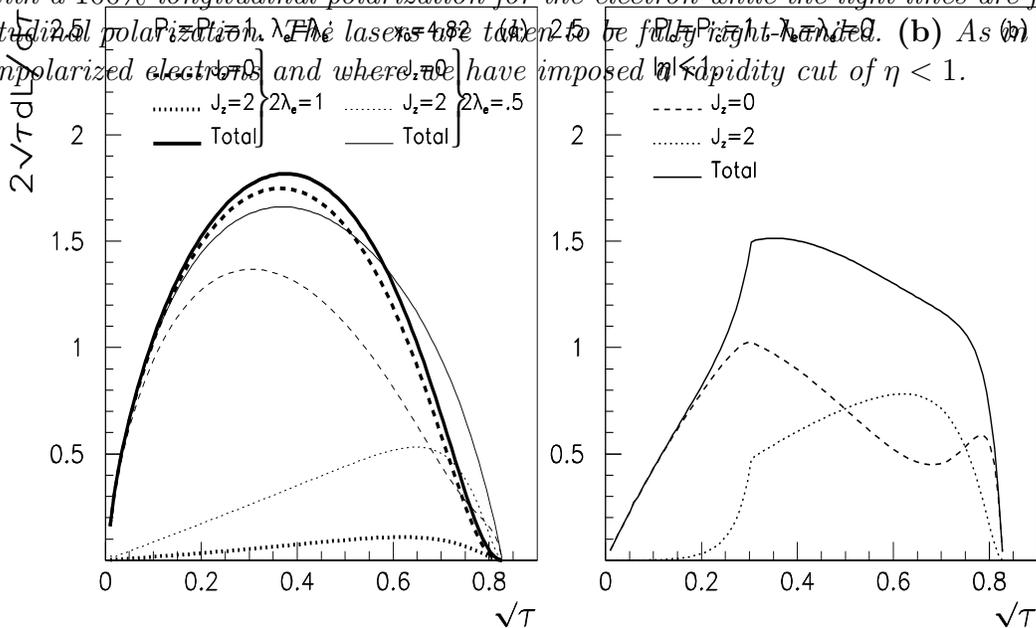}}
 \vspace*{-1.cm}
\end{center}
\end{figure*}

For the Higgs search, that is when we would like to keep an almost
constant value for the differential luminosity, the ``broad"
spectrum that favours the $J_Z =0$ is highly recommended. What is
very gratifying is that with $P_c=P_c'=2\lambdae=2\lambdae'=1$ the
whole spectrum is accounted for almost totally by the $J_Z=0$
spectrum(see Fig.~\ref{spectre34}a); the $J_Z=2$ contributes
slightly only at the higher end. This near purity of the $J_Z=0$
is not much degraded if the maximum mean helicity of the electron
is not achieved. We show on the same figure
(Fig.~\ref{spectre34}a) what happens when we change both
$2\lambda_e$ and $2\lambda_e'$ from $1$ to $.5$, keeping
$P_c=P_c'=1$. There
 is still a clear dominance of the $J_Z=0$ especially for the lower values
of the centre-of-mass energy. We would like to draw attention to
the fact that this effect, (increasing the $\frac{J_Z=0}{J_Z=2}$
ratio), can be further enhanced (when the maximal electron
polarization is not available) by imposing rapidity cuts.

Let us now be a bit more realistic and turn to effect of the
distance of conversion, keeping a blind eye on the soft electrons,
and the magnetic field.
\begin{figure*}[hbtp]
\caption{{\bf (a)} {\em The total luminosity spectrum with
$2\lambda_e P_c=2\lambdae' P_c'=-1$ (``peaked spectrum") for
different values of the conversion distance taking a spotsize
$\sigma_e=200$~nm.} {\bf (b)} {\em As in {\bf a} but with
$2\lambda_e P_c=2\lambda'_e P_c'=1$ (``broad spectrum").}}
\begin{center}
\vspace*{-1.5cm}
\mbox{\epsfxsize=14cm\epsfysize=9.cm\epsffile{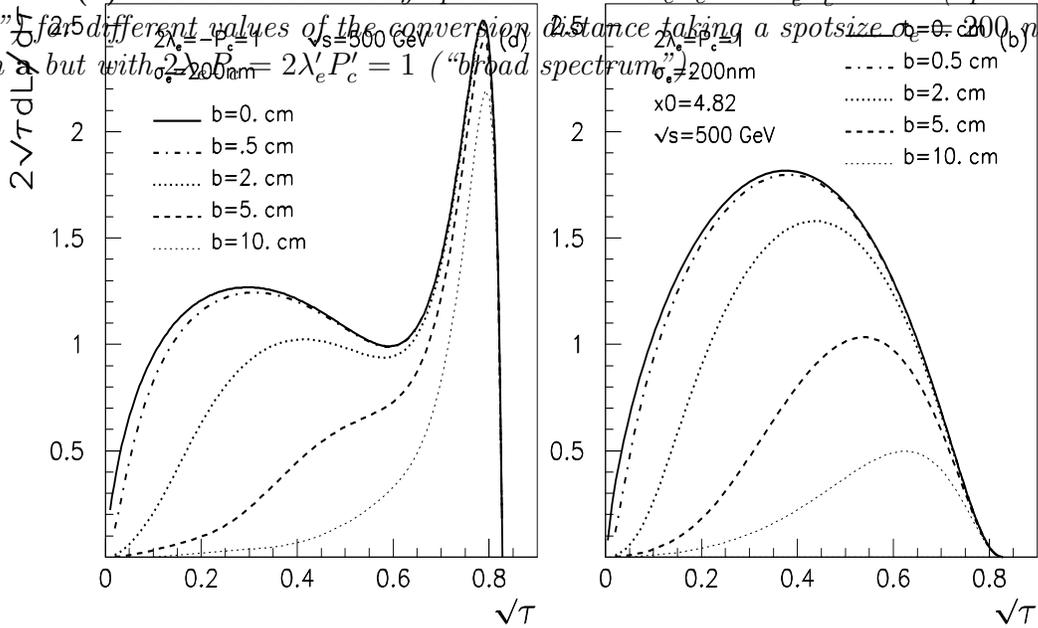}}
\vspace*{-1.cm}
\end{center}
\label{spectre56}
\end{figure*}

\noi As explained above, increasing the distance of conversion
filters the high energy modes and therefore the spectrum becomes
more monochromatic for large values of \gag centre-of-mass energy.
 For the peaked
spectrum, arrived at by having
 $2 \lambda_e P_c=2 \lambda'_eP_c'=-1$, the peaking
is dramatically enhanced for a large conversion distance $b=10$cm
($\rho_0\simeq 0.72$). This means for example that with a
conversion distance of $5$cm or $10$cm, there is almost no
luminosity below $\sqrt{\tau}<0.65$. This also means (see
Fig.~\ref{spectre56}a) that the spectrum is a purely $J_Z=0$
peaked spectrum. This is the most ideal situation to study a
$J_Z=0$ resonance if its mass falls in this energy range, {\it
i.e}, $ 0.7 < \sqrt{\tau_{res.}} < 0.82 $. The $J_Z=2$ component
that was present for the zero-distance of conversion is
effectively eliminated for large distances $b>5$cm. Note that in
this case if one could ``manage" with a conversion distance of
$0.5$cm then we almost recover the $b=0$ spectrum. \\ The
situation is not as bright for the broad spectrum case when the
interest is on small $\sqrt{\tau}$, like the search of an
intermediate-mass Higgs (IMH) at a $500$~GeV \epm. The nice
features that were unraveled in the last paragraph (an almost pure
$J_Z=0$ for small to moderate $\sqrt{\tau}$) are lost because the
luminosity in the energy range of the IMH peak formation is
totally negligible for conversion distances of order $\sim 5$cm or
higher (see Fig.~\ref{spectre56}b). If one could manage with a
conversion distance below $2$cm then we may hope to keep the nice
features of the ``broad" $J_Z=0$ scheme.

\begin{figure*}[hbtp]
\caption{\label{realisticspectrum}{\em Luminosity spectra with $2\lambda_e P_c=2\lambdae'
P_c'=-1$ based on a simulation using the TESLA parameters}\cite{ggTESLA}. The second spectrum
correspond to having a sweeping magnet. The example we show here is in fact based on one of the most
optimistic scenarios for the \gamgamt based on TESLA, the beams are assumed almost flat here!}
\begin{center}
\vspace*{2cm} \mbox{\epsfxsize=12cm\epsfysize=5.cm\epsffile{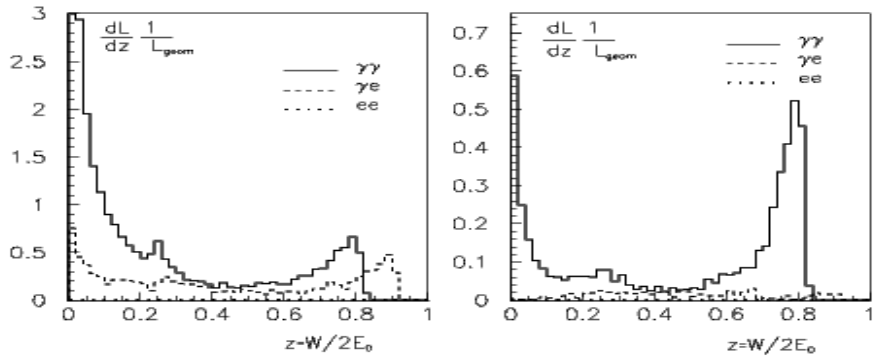}}
\vspace*{-1.cm}
\end{center}
\end{figure*}

This said, new simulations\cite{ggTESLA} have been conducted that have taken into account
the TESLA parameters for the electron beam and studied the spectra one obtains with the
option of deflecting the electron. It turns out that multiple rescattering has the effect
of considerably enhancing the lower end of the spectrum. In order not to end up with too
small a \gamgamt luminosity for large $\sqrt{\tau}$ it is advisable to deflect the
electrons. Still as can be seen from Fig.~\ref{realisticspectrum} which adopts some
optimised parameters for the TESLA design, although the deflection scheme reduces the
\egamt noise, the peak luminosity for the energetic end of the spectrum is about 5 times
lower than what we obtained with the idealistic distributions. Therefore most of the
studies (for reviews see\cite{Parisgg,Brodskygg,ggTESLA}) that have been performed for
this type of collider should be critically re-analyzed.

\subsection{Typical cross section: the would-be-backgrounds}
\begin{figure*}[htbp]
\vspace*{-0.5cm} \caption{\label{wfactory}{\em Typical sizes of
cross sections for weak boson production at the linear colliders
in the different modes of the machine. No convolution with
luminosity spectra have been applied.}}
\centerline{\resizebox*{7.5cm}{14cm}{\includegraphics{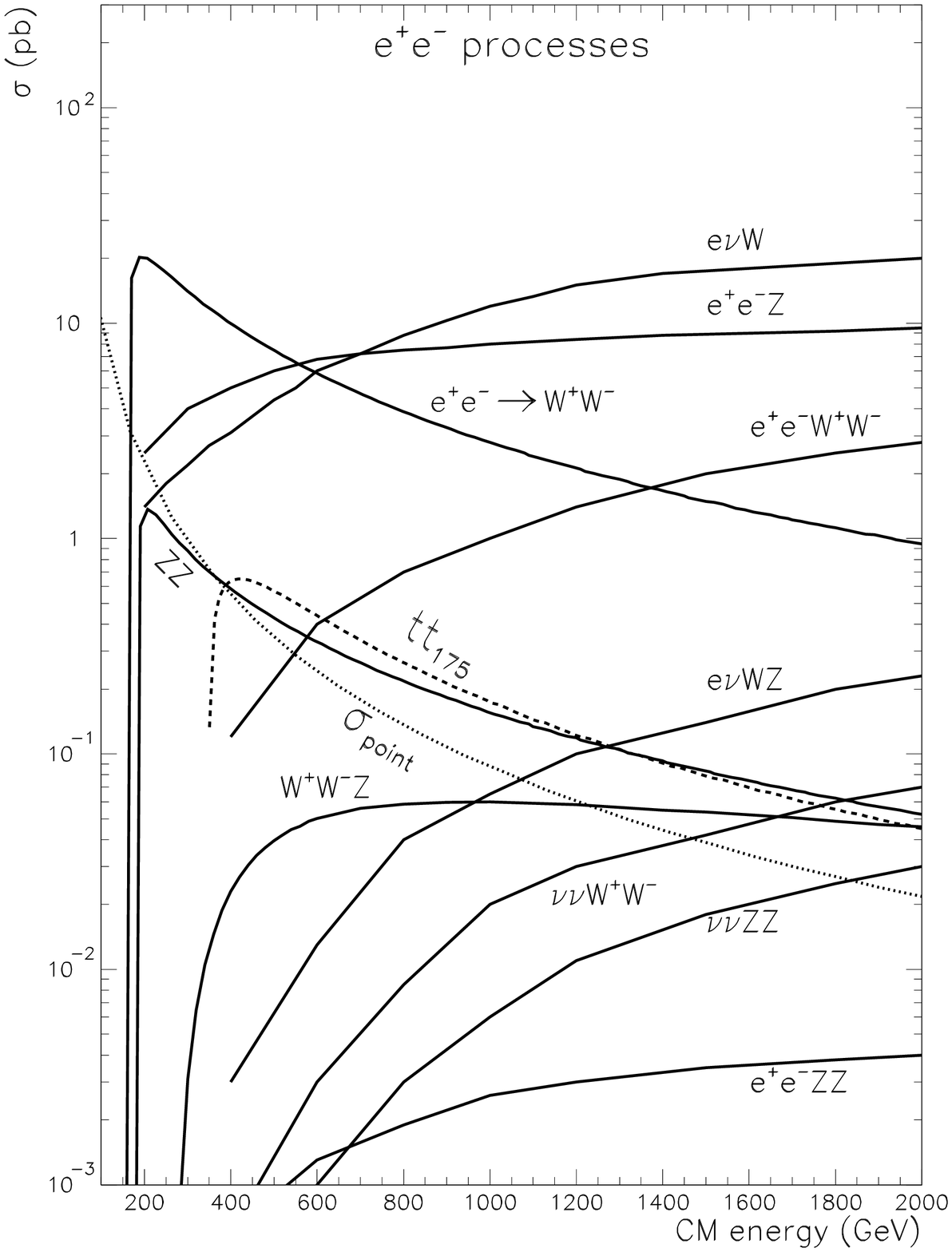}}
\resizebox*{7.5cm}{14.5cm}{\includegraphics{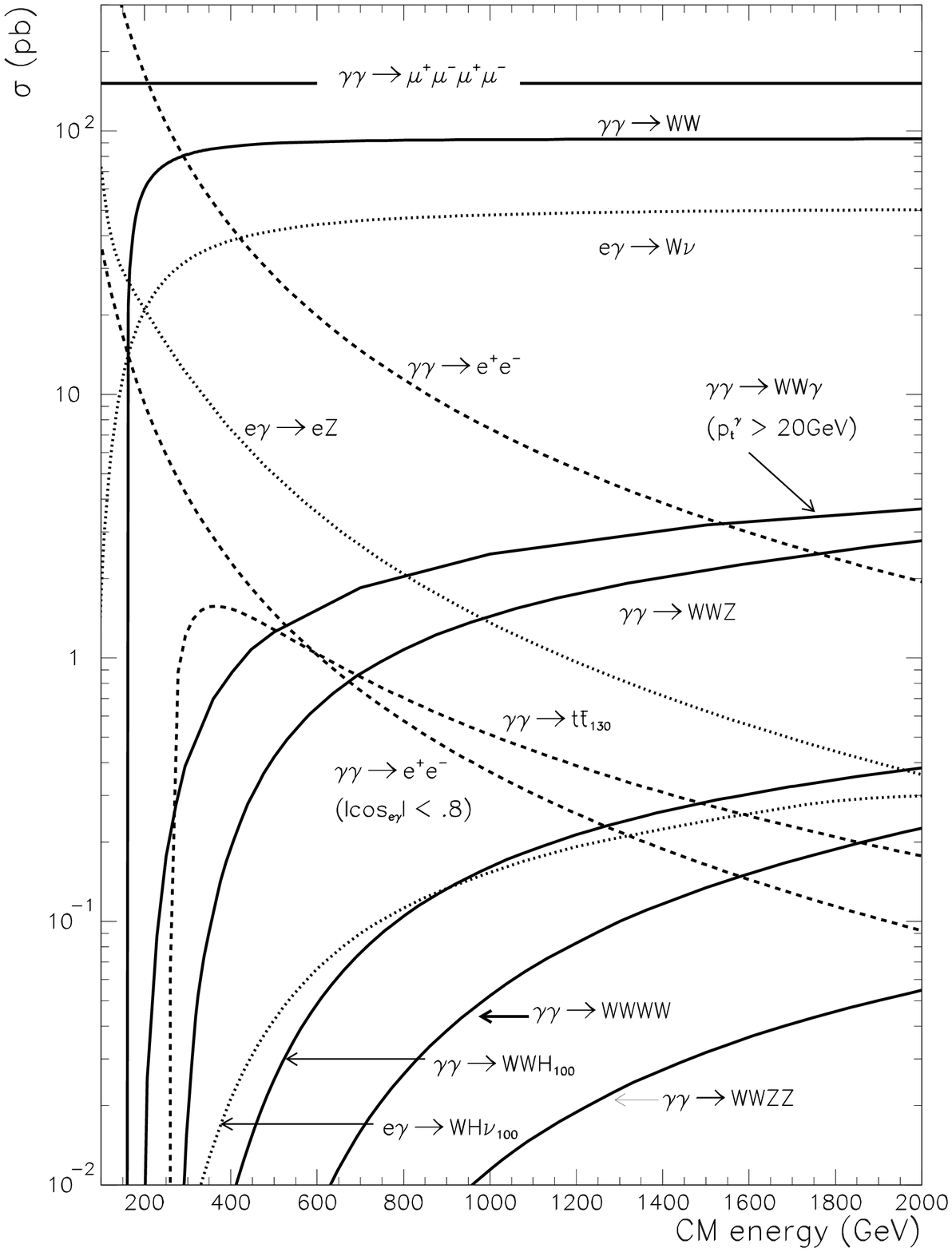}}}
\end{figure*}
Production of new particles proceeding essentially through the
s-channel in \epemt have cross sections of few $fb$. The main
backgrounds at the linear collider will be dominated by $W$
processes. Indeed as can be seen from Fig.~\ref{wfactory} cross
sections for production of $W$'s and $Z$'s either in the \epemt
mode or the \gamgamt (or for that matter the \egamt mode) can
reach a few picobarn ( we can see that the point cross section,
$\sigma_{\rm point}$) is buried in the electroweak background).
Fortunately the bulk of these processes is rather in the forward
region, moreover they are quite sensitive to the beam
polarization.

Especially in the \gamgamt mode, hadronic processes constitute a
formidable background.
\begin{figure*}[htbp]
\caption{\label{gghadronic}{\em Direct and resolved photon
contributions to bottom and charm production at \gamgamt collider
obtained from \epemt at 300GeV (figure on the left) and 500GeV (on
the right). The signal from a 120GeV SM Higgs as well as the
additional induced $Z$ background is shown. For the resonance a
smearing of 5GeV is applied. The photon spectrum is the idealised
with $\lambda_e=.9$ $P_c=1$.}}
\begin{center}
\vspace*{0.8cm} \hspace*{.2cm} \mbox{
\mbox{\epsfxsize=7.5cm\epsfysize=10cm\epsffile{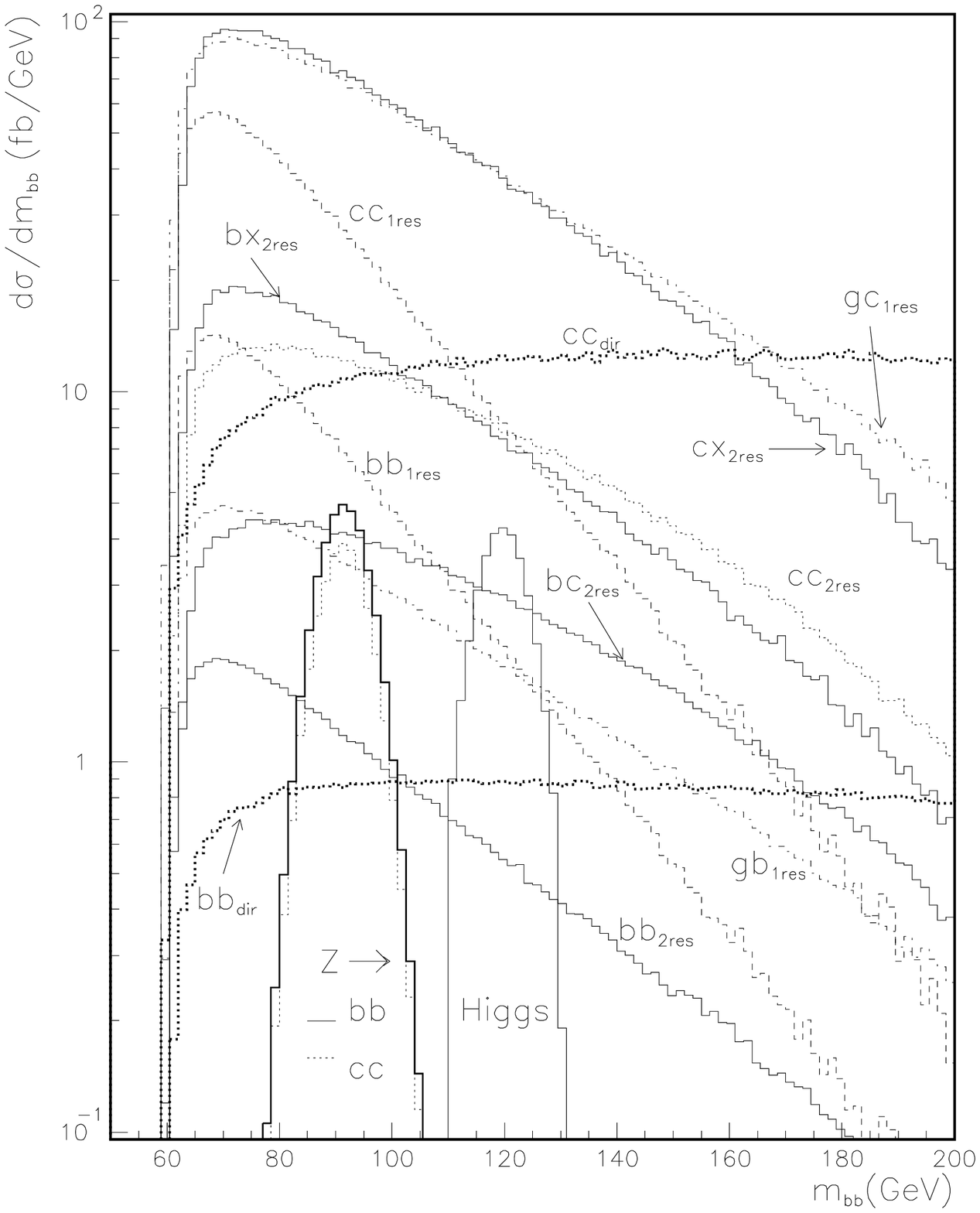}}
\hspace*{0.5cm}
\mbox{\epsfxsize=7.5cm\epsfysize=10cm\epsffile{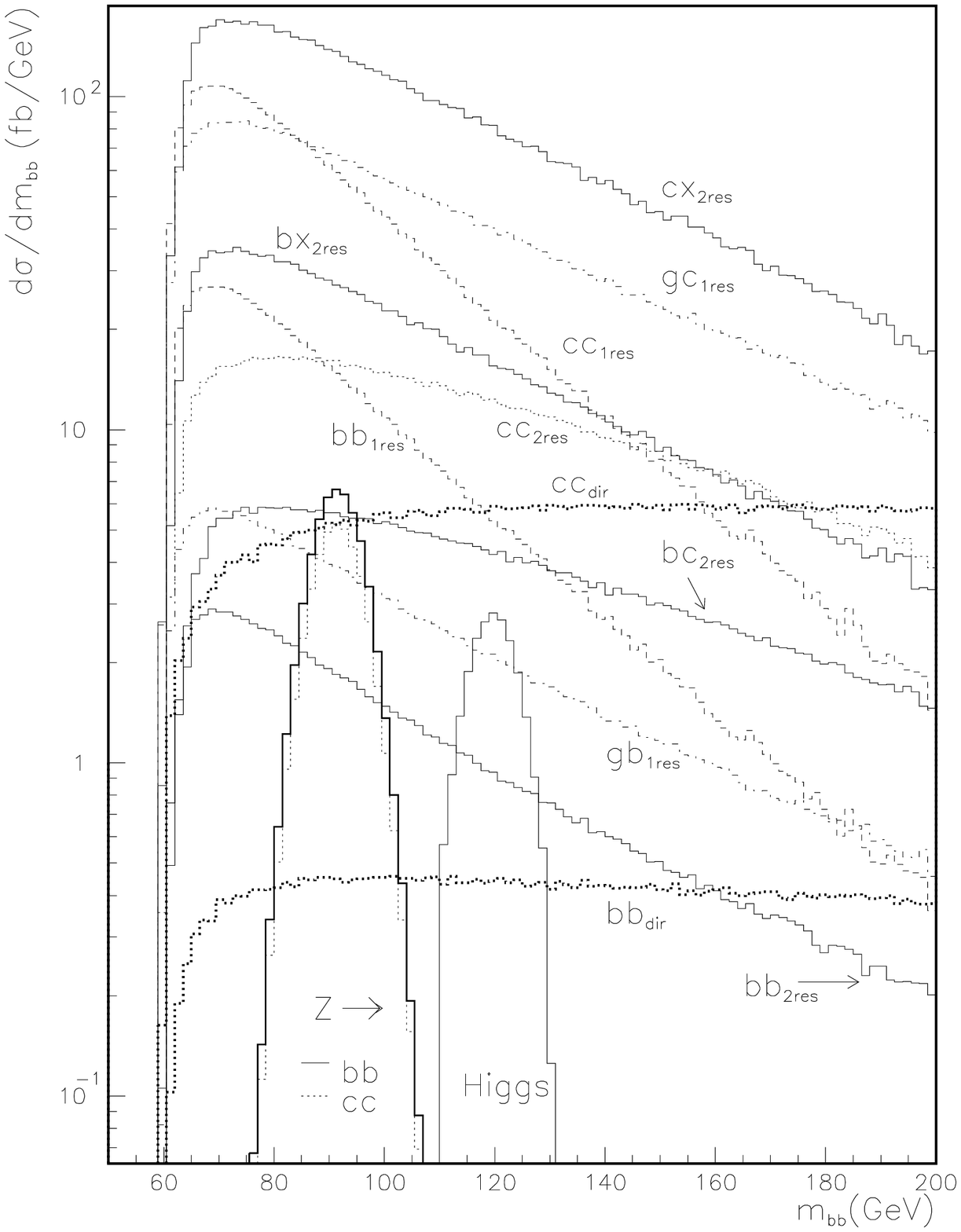}}}
\end{center}
\end{figure*}
If one has a wide spectrum with invariant photon masses that
extends beyond 300GeV, the bulk of these hadronic events are
induced through resolved photon contributions (splitting into
quarks and gluons,..). What is even more dramatic is that the
initial photon polarisation will not be transferred to these
constituents, thereby we loose the control of reducing
backgrounds. This is especially serious for searches of the
intermediate mass Higgs, IMH ( which decays into a $b\bar b$ pair)
as a resonance\cite{Halzen}. The latter can be produced by
selecting $J_Z=0$ set-up which has the advantage of drastically
reducing the direct contribution $\gamgam \ra q \bar q$.
Figure~\ref{gghadronic}\cite{Noushiggsres} shows how the Higgs
resonance gets buried under the background, if one chooses a the
laser set-up so that one has a broad spectrum. These cross
sections ought to be kept in mind when we seek New Physics.
Although in the previous example it has been shown how to salvage
the situation, it is worth observing that for higher Higgs masses
the background (into $t \bar t$) reduces quite a bit. Also if one
wants to make precision studies of the IMH\cite{Borden}, then it
is best to tune the machine so that the spectrum does peaks around
the Higgs mass, this would though preclude a host of other
studies.

\section{The main issue: the mass problem and symmetry breaking}
Considering that the LHC is a certainty, the fact that it has an
energy reach greater than the linear collider and that the latter
will certainly not be built before the LHC it is important to ask
why one needs a linear collider. \epemt machines have always had
the advantage of more than making it up for their lack of phase
space by offering a clean environment which is conducive to
precision measurements. For instance, we all know how difficult it
is to discover the IMH at the LHC\cite{HiggsLHC}. This is even
more frustrating since despite the fact that \susy does predict an
IMH even if all other particles can be too heavy one will most
probably have to await the high luminosity option of the LHC and
combine the data of ATLAS and CMS in order to unravel it. In
contrast, a 500GeV centre-of-mass energy \epemt collider, with a
very humble luminosity $10fb^{-1}$, will discover the same Higgs
in a matter of weeks (even days). This has far-reaching
consequences: if no Higgs is seen even in the first phase of the
LC the \susy scenario will be out! As we will argue, even if one
discovers new particles at the LHC, the main issue will be to
better understand its origin. For instance even if \susy is
discovered at the LHC one would like to understand the mechanism
of its breaking and reconstruct the vast array of the parameter
space that plague the current phenomenological description of
\susy. As a matter of fact, probably the main raison d'\^etre of
the LC will be the understanding of symmetry breaking, a mechanism
one has had till now little insight.\\
It should be remembered
that the stunning success of the standard model, \sm, is based on
the fact that the model reconciles the gauge symmetry principle
(in its non-Abelian form) with the apparent breaking of this
symmetry by giving masses to the gauge bosons and fermions. The
gauge symmetry principle has now been tested at the per-mil level
through the universality of various gauge couplings. Even the
(most evident) non-Abelian vertices ($WW\gamma, WWZ, WWWW,...$)
are now badly required by precision data through their effects at
the quantum level. However, to be fair these tests concern
essentially the transverse polarizations of the vector bosons.
Apart from the presence of the mass terms, one knows very little
about the longitudinal vector components of the bosons (this is in
a sense the physics of the Goldstones) and how exactly the
left-handed and right-handed components of fermions interact with
each other (this concerns essentially the top). These aspects are
intimately related, in the \sm description, to the Higgs
mechanism. Not only the particle it predicts is still missing,
though the latest global fits tend to indicate a not too heavy
Higgs, but this particle does pose some very uncomfortable
naturality problems which cast doubt on the whole thing and
strongly suggest some alternative scenario that the planned
colliders seek to uncover. In a nutshell, it is best to think of
the naturality argument as being intimately related to the fact
that there is no symmetry associated to the mass of an elementary
spin-less particle. Chiral symmetry prevents fermions masses while
gauge symmetry prevents vector boson masses with the consequence
that radiative corrections to these masses are only
logarithmically divergent (prior to renormalisation of course).
Lack of a symmetry means that there is no reason why the mass of a
scalar should be kept small and hence the infamous quadratic
divergence.
 To remedy this,  one option is  to
make do without an elementary scalar. One then inevitably has to
deal with a strongly interacting phase of the weak interactions
with the formation of condensates and bound-states and thus little
calculabilty and much reduced predictivity. Or one tries to
implement a symmetry. The most popular and attractive option is
supersymmetry where a scalar and a fermion become the {\em avatar}
of the same multiplet, and thus the scalar inherit the chiral
symmetry and is protected. But then again, one knows that this
symmetry is far from being perfect: the associated scalar and
fermion ought to have the same mass. Then even if hints of \susy
are revealed or super-particles discovered the pressing question
is how is supersymmetry broken. Lacking a fundamental theory for
this breaking one has to parameterize it by a large number of
parameters. Again this can be addressed through precision tests
which are best conducted in \epemt because of its cleanliness and
also because of the availability of polarisation. The latter is a
wonderful tool to study symmetry breaking, \sb. \sb can be seen as
due to the mixing and interaction of states with different quantum
numbers: left and right for fermions which have different isospin
numbers with the consequence that their super-partners inherit
also the same quantum numbers. For bosons the presence of the
Golstones and the transverse modes means that in the \susy version
one has to deal with the mixing of gauginos and Higgsinos. Thus
controlling the polarisation of a \sm state can help reach its
\susy partner which is some component of a physical \susy state.

The remaining plan of the talk evolves around three main
manifestations of the Higgs potential that describe the three main
possibilities describing electroweak \sb. 1) In the no Higgs
scenario (condensates, etc..) the scalar potential does not
appear, or at least is not described in terms of fundamental
fields. We will then see how the physics of the Goldstones may
shed light on the New Physics. 2)In the standard model
description, the puzzle in the Higgs potential is the negative
mass squared (associated though to the Higgs doublet) and the fact
that self-interaction of the Higgs $\lambda$ (which determines the
mass of the Higgs) is not fixed
\beqn
V=\lambda
\biggl(\Phi^\dagger \Phi- \frac{|\mu|^2}{2\lambda}\biggr)^2\;+\;
V_0=-|\mu|^2 \Phi^\dagger \Phi+ \lambda |\Phi^\dagger \Phi|^2 +...
\eeqn
If this is all we have, understanding of the properties of
the Higgs will be the bread and butter of the LC. 3) In \susy, the
situation is better. $\lambda$ acquires the status of a gauge
coupling, with the dramatic effect that the mass of the lightest
Higgs is bounded, at tree-level, to be less than $M_Z$. However
the negative square mass that drives symmetry breaking is still
ad-hoc. There are tantalizing scenarios which embed \susy in a
grand scheme whereby the origin of the "negative square mass" is
dynamical. One such scenario is the popular minimal SUGRA model
where all scalar masses are universal (with a "positive square
mass") at the GUT scale. The heavy top drives one of the "Higgs
masses" negative as one runs down to lower energies.
 Supersymmetry breaking though
is still obscure and, in fact, the issue is relegated to a hidden
sector, although such schemes do provide sum rules like the
equalities of scalar masses, and gauginos masses at the high
scale. It is these kinds of sum rules and implementations that one
hopes precision measurements at the LC can unravel thus giving us
indirect probes of physics much beyond the TeV scale.

\section{No Higgs scenarios and the $W$ couplings}
Current global fits to the electroweak data tend to prefer a light
Higgs mass with $M_H < 420GEV \;95\%CL$. However in face of the
discrepancy between the SLD and LEP data on the effective weak
mixing angle, it has been suggested to be careful when quoting
these kinds of limits. Taking the LEP data alone weakens the bound
to $M_H < 700-800GEV \;95\%CL$\cite{SchildknechtGI97}. So
especially when planning for the future when should still
entertain a scenario where a Higgs is too heavy or simply not
there. This said one has to implement the gauge symmetry without
the scalar potential. As stressed above gauge symmetry is now
sacrosanct. Already in 1994, fitting the electroweak data without
the non-Abelian gauge vertices gave more than $7\sigma$
departure\cite{SchildknechtGI94}. To implement gauge invariance an
yet make do without the Higgs one retorts to a non-linear
realization of symmetry breaking. To have more predictivity in
this approach one can appeal to another well confirmed symmetry:
the global SU(2) custodial symmetry which is the most natural
explanation for the fact that once the top-bottom splitting has
been taken into account the $\rho$ parameter is essentially unity.
One therefore may assemble the triplet of Goldstone Bosons
$\omega_i$ into the matrix-field $\Sigma=exp(\frac{i
\omega_{\alpha} \tau^{\alpha}}{v}) \;\;\; (v=246GeV)$ and define
the covariant derivative (through which gauge invariance will be
maintained) $\cd_\mu=\partial_\mu \Sigma + \frac{i}{2} \left( g
\frac{\tau^i}{2} W_{\mu}^i \Sigma - g'B_\mu \Sigma \tau_3
\right)$. The weak bosons mass term writes \beqn \label{masscov}
\cl_M=\frac{v^2}{4} \tr(\cd^\mu \Sigma^\dagger \cd_\mu \Sigma)
\;\;\; ; \;\; \Sigma=exp(\frac{i \omega_{\alpha}
\tau^{\alpha}}{v}) \ra M_W^2 W^+_\mu W^{-\mu}+\frac{1}{2} M_Z^2
Z_\mu Z^\mu \eeqn In this so-called non-linear realization of \sb
the mass term for the $W$ and $Z$ is formally recovered by going
to the physical ``frame" (gauge) where all Goldstones disappear,
{\it i.e.}, $\Sigma \ra${\bf 1}. With this description the model
is not renormalizable, however it can be made finite by the
introduction of a cut-off which exhibits the same dependence as
that of the Higgs mass in loop effects. This cut-off represents
the on-set of New Physics. We expect that before this scale is
reached, which is the case with the first stage LC, the effect of
the New Physics will contribute to a few operators that are not
described by the minimal \sm. In fact the above mass operator
Eq.\ref{masscov} should be considered as the leading (lowest)
operator in an energy expansion. With the custodial symmetry and
the requirements of gauge invariance there are few other operators
related to the Goldstone sector that we may write. They give
contributions to the self-couplings of the weak vector bosons,
especially their longitudinal parts. They can be probed both at
the LHC and LC in a variety of weak boson production and
scattering. For the 500GeV LC the most important operators are
given by ${{\cal L}}_{9L,R}$ (for details, see\cite{Morioka})
\beqn
{{\cal L}}_{9R}&=&-i g'
\frac{L_{9R}}{16 \pi^2} \tr ( \B^{\mu \nu}\cd_{\mu}
\Sigma^{\dagger} \cd_{\nu} \Sigma ) \nonumber \\ {{\cal
L}}_{9L}&=&-i g \frac{L_{9L}}{16 \pi^2} \tr ( \W^{\mu
\nu}\cd_{\mu} \Sigma \cd_{\nu} \Sigma^{\dagger} )
\eeqn

If probed efficiently these operators can tell us something about
the dynamics of the Goldstones. Noting that the Godlstones are
contained in the covariant derivative, $\cd_\mu$, whereas the
transverse are essentially described by the field-strengths and
that the \sm processes are dominated by the transverse modes, one
should select the longitudinals. Thus for the above operators, one
should maximise their effects by having the Goldstones
contributing in the final state. Thus $f\bar f \ra W^+ W^-$ seems
the most appropriate. However in the $pp$ environment this has
either a huge hadronic background or can not be fully
reconstructed because of the two missing neutrinos. Reverting to
$pp \ra W\gamma, WZ$ means that the first operator will be very
poorly probed at the LHC. In \epemt one can disentangle between
the two operators most easily through initial polarisation in
\eewwt, since the former couples only to the hypercharge component
and is thus enhanced if right-handed electrons are chosen. Both
can also be efficiently probed in the \gamgamt mode. In both
\gamgamt and \epemt to optimise the limits by accessing a maximum
of distributions in the kinematical variables of the decays, which
is somehow reconstructing the longitudinal and transerve
polarisations. To that effect one sees that by writing the $WW$
final state in terms of the four-fermions, all the helicity
amplitudes are accessed.

\beqn
\label{fullspincorr} & & \mbox{}\frac{ {\rm
d}\sigma(\gamma(\lambda_1)\gamma(\lambda_2) \ra W^+W^-\ra f_1 \bar
f_2 f_3 \bar f_4)}
     { {\rm d}\cos \theta \;\;{\rm d}\cos \theta_-^{*} \;\;{\rm d}\phi_-^{*}\;\;
      {\rm d}\cos \theta_+^{*} \;\;{\rm d}\phi_+^{*} }=Br^{f_1 \bar f_2}_W Br^{f_3 \bar f_4}_W
\frac{\beta}{32\pi s}
\left( \frac{3}{8 \pi}\right)^2 \hfill \nonumber \\ \lefteqn{
\sum_{\lambda_- \lambda_+ \lambda'_- \lambda'_+}
 {\cal M}_{\lambda_1,\lambda_2; \lambda_-\lambda_+} (s,\cos \theta)\;
{\cal M}_{\lambda_1,\lambda_2; \lambda'_-\lambda'_+}^{*} (s,\cos
\theta) \; D_{\lambda_- \lambda'_-} (\theta_-^{*} ,\phi_-^{*}) \;
D_{\lambda_+ \lambda'_+} (\pi-\theta_+^{*}, \phi_+^{*}+\pi)}
\nonumber \\ \lefteqn{ \equiv \frac{{\rm
d}\sigma(\gamma(\lambda_1)\gamma(\lambda_2) \ra W^+W^-)}{{\rm
d}\cos \theta} \left( \frac{3}{8 \pi}\right)^2 Br^{f_1 \bar f_2}_W
Br^{f_3 \bar f_4}_W } \nonumber \\ \lefteqn{ \sum_{\lambda_-
\lambda_+ \lambda'_- \lambda'_+} \rho_{\lambda_- \lambda_+
\lambda'_- \lambda'_+}^{\lambda_1,\lambda_2}\; D_{\lambda_-
\lambda'_-} (\theta_-^{*} ,\phi_-^{*}) \; D_{\lambda_+ \lambda'_+}
(\pi-\theta_+^{*}, \phi_+^{*}+\pi) } \nonumber \\ &{\rm with}&
\hspace*{.6cm}\rho_{\lambda_- \lambda_+ \lambda'_-
\lambda'_+}^{\lambda_1,\lambda_2}(s, \cos\theta)= \frac{ {\cal
M}_{\lambda_1,\lambda_2; \lambda_-\lambda_+} (s,\cos \theta)\;
{\cal M}_{\lambda_1,\lambda_2; \lambda'_-\lambda'_+}^{*} (s,\cos
\theta)} { \sum_{\lambda_- \lambda_+} |{\cal
M}_{\lambda_1,\lambda_2; \lambda_-\lambda_+}(s,\cos \theta)|^2},
\eeqn \noindent where $\theta$ is the scattering angle of the
$W^-$ and $\rho$ is the density matrix. A maximum likelihood
fitting procedure exploiting all the decay angles permits to put
very tight bounds on the $L_9$ parameters, whereas in the $pp$
environment on relies on a much reduced set of variables and
sometimes only on the counting rate.

\begin{figure*}[hbtp]
\caption{\label{l9bounds}{\em Limits on ($L_{9L}-L_{9R}$) in
\epemt including ISR and beam polarisation.}}
\begin{center}
\vspace{-1.5cm}
\mbox{\epsfxsize=12cm\epsfysize=12cm\epsffile{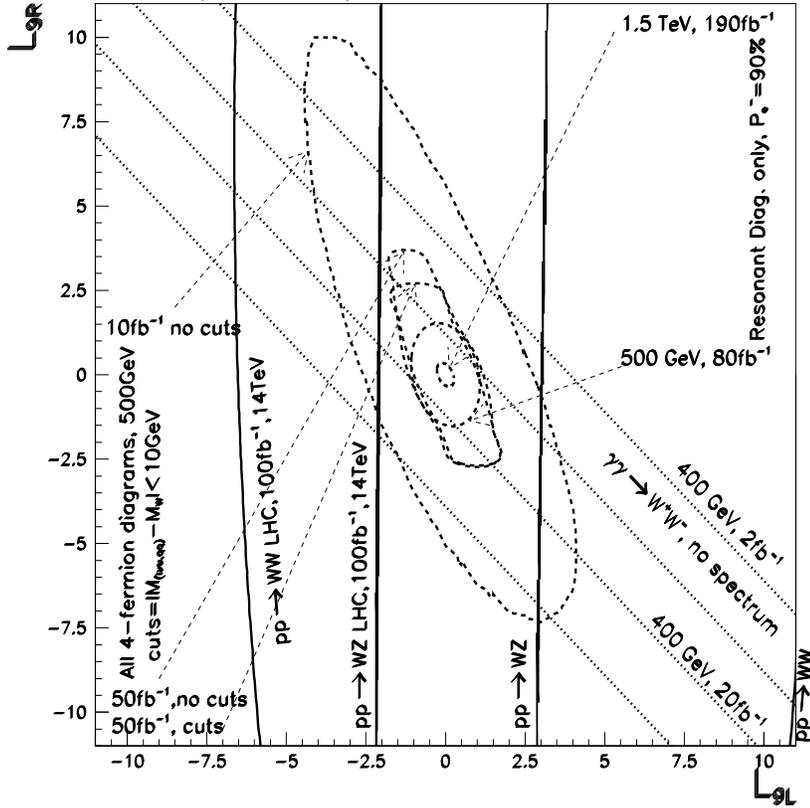}} \vspace*{-1cm}
\end{center}
\end{figure*}
\vspace*{0.5cm}

These observations are well rendered by Fig.~\ref{l9bounds} which
is a compendium of studies at both \epemt, \gamgamt and the
LHC\cite{Morioka}. One sees that already with a $500$GeV \epemt
collider combined with a good integrated luminosity of about
$80fb^{-1}$ one can reach a precision, on the parameters that
probe \sb in the genuine tri-linear $WWV$ couplings, of the same
order as what we can be achieved with LEP1 on the two-point
vertices. To reach higher precision and critically probe \sb one
needs to go to $TeV$ machines. In fact, at an effective $WW$
invariant masses of order the TeV, \sb (especially in
scalar-dominated models) is best probed through the genuine
quartic couplings in $WW$ scattering or even perhaps in $WWZ, ZZZ$
production (that are poorly constrained at $500$GeV). LHC could
also address this particular issue but one needs dedicated careful
simulations to see whether any signal could be extracted in the
$pp$ environment. In this regime there is also the fascinating
aspect of $W$ interaction that I have not discussed and which is
the appearance of strong resonances and the study of $WW$
scattering. This would reveal another alternative to the \sm
description of the scalar sector but can only be studied at a
1.5-2TeV LC or better with a 4TeV muon collider. \\

\section{Properties of the Higgs}
Unlike the situation at the LHC, the Higgs can be very easily
discovered at the LC\cite{DesyNLC} almost up to the kinematical
limit through $\epem \ra ZH$ and and $WW(ZZ)$ fusions $\epem \ra
\nu_e (e^+) \bar{\nu}_e (e^-) H$. The point is whether one can
learn more from this. If one looks at the Higgs interactions in
the \sm, \beqn {{\cal L}}_{H,M}=(D_\mu \Phi)^\dagger (D_\mu \Phi)+
{{\cal L}}_{Yukawa} -V \eeqn the generalized kinetic term contains
the mass terms of the $W/Z$ bosons but also the couplings of the
Higgs to the weak vector bosons. The latter trigger the main Higgs
production mechanisms in \epemt. As in the previous section one
can check whether there are higher order operators that modify
these couplings as well as the tri-linear $WWV$ couplings. More
interesting, and paving the way to the \susy tests, is to check
whether there is only one Higgs doublet which is giving mass to
the weak bosons. If this is the case then only one {\it v.e.v.} is
involved and thus, for instance, the $ZZH$ coupling is completely
specified by the gauge couplings and the $Z$ mass. If there were
more than one doublet (and hence more than one Higgs) the $ZZH_i$
couplings will depend on ratios of v.e.v and would therefore be
smaller than if there were only one Higgs. Therefore, by precisely
measuring the cross section of Higgs production one could in
principle infer the presence of another Higgs. Exactly the same
conclusion applies to the Yukawa couplings. Most important is the
measurement of $H\ra b \bar{b}$, which for example in \susy
depends crucially on $tg\beta$. \\ At the LC, it is possible to
optimize the running conditions by lowering $\sqrt{s}$ if
necessary. For example for $M_H<200GeV$ one could choose
$\sqrt{s}\simeq 300GeV$ to maximize $\epem \ra ZH$. The other
advantage over $pp$ colliders is that one can, again as in the
previous section, use all topologies (all $Z$ decays may be
used!). In these conditions with quite modest luminosities
(~20$fb^{-1}$) it is found\cite{Kooten} that one can measure the
mass of the Higgs at the per-mil level. The $ZZH$ coupling may be
measured at the 5per-cent level. Notice that such a precision does
not seem to be enough to discriminate the \sm Higgs with a minimal
\susy Higgs. Indeed if a large deviation in this coupling is found
$hA$ production should have been observed, otherwise such a
precision will not reveal the indirect presence of an extra Higgs.
A slightly more hopeful conclusion holds for the $h\ra b\bar{b}$
where  if the mass of the pseudo-scalar Higgs is above $400GeV$,
the $Br(h\ra b\bar{b})$ can not be larger than about $7\%$.
Simulations based on the the $ZH$ final state at $500GeV$ have
found that for $M_h=120GeV$ the branching ratio can be measured at
$7\%$ but only at $12\%$ for $M_h=140GeV$. More thorough
simulations should be performed on this coupling. For the other
couplings, branching fractions are measured with a much worse
precision. Apart from this, we note that some nice checks on the
spin and parity of the Higgs can be performed\cite{DesyNLC}. First
in \epemt, from the angular distribution of the Higgs or the
reconstructed $Z$, one could tell whether the parity of the
particle is odd or even. Take $\epem \ra ZH$, with $x$ denoting
the cosine of the scattering angle, one has

\beqn \frac{1}{\sigma} \frac{{\rm d}\sigma}{{\rm d}x}&=& (1+x^2)\;+\;\frac{s}{8\mzz}
\biggl(1+\frac{\mzz}{s}-\frac{M_h^2}{s}\biggr)^2 (1-x^2) \propto
(1-x)^2 +{{\cal O}}(M_Z^2/s) \ra {\rm parity} \;{\rm even}
\nonumber \\ \frac{1}{\sigma} \frac{{\rm d}\sigma}{{\rm
d}x}&\propto& (1+x^2)\ra {\rm parity} \;{\rm odd} \eeqn This may,
nonetheless, prove to be an academic exercise since a $ZZ{\rm
scalar}$ requires a parity-even scalar. The \gamgamt mode can help
in many ways as far as the Higgs is concerned. First, by choosing
the polarizations such that the colliding photons are in a $J_Z=0$
state(photons with the same helicity) producing a particle as a
resonance gives its spin unambiguously. Moreover, to check for CP
violation in case the scalar is an admixture of a CP even and a CP
odd state, one should look for an asymmetry between the two
$J_Z=0$ configurations, that is depending on whether both photons
are right-handed or both are left-handed. Another trick for the
parity measurement is to invoke linear polarisation. A parity even
scalar, $O^+$ couples as $F_{\mu \nu} F^{\mu \nu} O^+$ and thereby
the two photon polarizations are parallel whereas for a parity odd
this combination is not possible\cite{DesyNLC}. One can also,
through the measurement of the cross section $\gamma \gamma \ra H
\ra b\bar{b}$, extract the $\Gamma (H\ra \gamma \gamma)$ {\it
width} assuming the branching ratio into b's has been measured in
the \epemt mode. A simulation has shown that this width can be
measured at $6\%$\cite{Kooten}. Note that for these precision
measurements to be possible in the \gamgamt mode on needs to
choose a peaked set-up at the Higgs peak, otherwise the $b\bar{b}$
background is killing for a light Higgs mass. For the heavier
neutral Higgses on the other hand, where the resolved photon
contributions are much smaller, one can use the maximum energy
possible in the \gamgamt mode in order to access the largest mass
as a resonance. This gives a wider range than in the \epemt mode.
Another proposal which needs more investigation, especially if no
direct sign of New Physics has been observed, is to retrieve the
LEP1/SLC data, or even better to run at the $Z$ peak with the LC
luminosity and polarisation. One can then input the Higgs mass,
the top mass which in passing can be measured with a precision of
.2GeV (this is almost a ten-fold better than at the LHC ) as well
as the measurement of $M_W$ which can be improved at the LC
($\Delta M_W=15MeV$). One thus have at hand some super precision
observables to infer some high-scale physics. Other interesting
tests concern the self-couplings of the Higgs. Within \susy these
couplings are essentially gauge couplings and thus these studies
are not as motivated as if a non-susy scalar has been discovered.
Unfortunately, one needs to go to high Higgs masses and energies
to probe anything useful\cite{ChopinHiggs}.

\section{SUSY and SUSY breaking}
If \susy is at work it will be a matter of days for the LC to
discover the lightest \susy Higgs. Else \susy will be shown not to
be the solution to the hierarchy problem. We have just discussed
the kind of checks that may be performed if only the lightest
Higgs were discovered and to what extent and conditions one might
infer from the precision measurements that it is actually a \susy
Higgs that one has discovered. On the optimistic side one might be
lucky and discover more than one Higgs if not all of them. This
occurs if the pseudo-scalar Higgs has a mass below $200-250GeV$ at
the 500GeV LC. As concerns the other \susy particles it is worth
putting the LHC in the picture. Indeed if \susy is at work, this
would mean that even if the LHC has had great difficulty cornering
a Higgs, it should have no problem producing plenty of coloured
\susy particles (gluinos and squarks). Many studies have  shown
that the LHC can cover a mass range for these particles up to
2TeV!\cite{SusyLHC}. If these are not produced we would be very
uncomfortable with \susy, since a new naturality problem creeps
in. LHC has also a good chance to see charginos and some
neutralinos, as well as sleptons. The problem is that all the
kinematically accessible particles will be accessed at once. The
heavier ones cascading into the lighter ones which will in turn
cascade into even lighter ones ....thus creating a very blurred
and  confusing picture. At least if one knew the \susy spectrum
and the \susy parameters one can reconstruct the original picture.
But we will not, and if one takes an unbiased attitude one will
have the formidable task to measure a large number of parameters.
On the other hand, once \susy is discovered it is exactly this,
measuring the \susy parameters, that will be a priority. This is
because one expects these parameters not to be completely
haphazard but show some simple structure that betrays some common
origin. Due to the nature of the supersymmetry transformations
this may even tie this model with gravity. There is also
circumstantial evidence that the unification of the gauge
couplings occurs within \susy. It is then utterly crucial to test
whether the unification of other parameters occurs as well. The
answer to these questions gives an information on physics at scale
orders of magnitude from the present ones, unreachable by any
collider. One has then to retort to ingenuity to extract this
information from the upcoming colliders.

Some simulations for the extraction of parameters has been
attempted for the LHC\cite{SusyLHC}. However it is important to
stress that these checks were done with the assumption of an
underlying model, minimal SUGRA that contains only a few
parameters. Although the parameters are extracted with a good
precision it must be remembered that these studies only confirm
whether a specific model is at work. The situation with
supersymmetry breaking may prove to be more complicated, so
ideally one would like to measure the parameters with no a priori
assumption about the model. This will, probably, not be possible
at the LHC.
\\

\begin{figure*}[htbp]
\caption{\label{selectron}{\em Same parameters as in the previous figure for the
scattering angle of the right-handed selectron\cite{JapanSusy}.}}
\begin{center}
\vspace{-1cm}
\mbox{\epsfxsize=9cm\epsfysize=9cm\epsffile{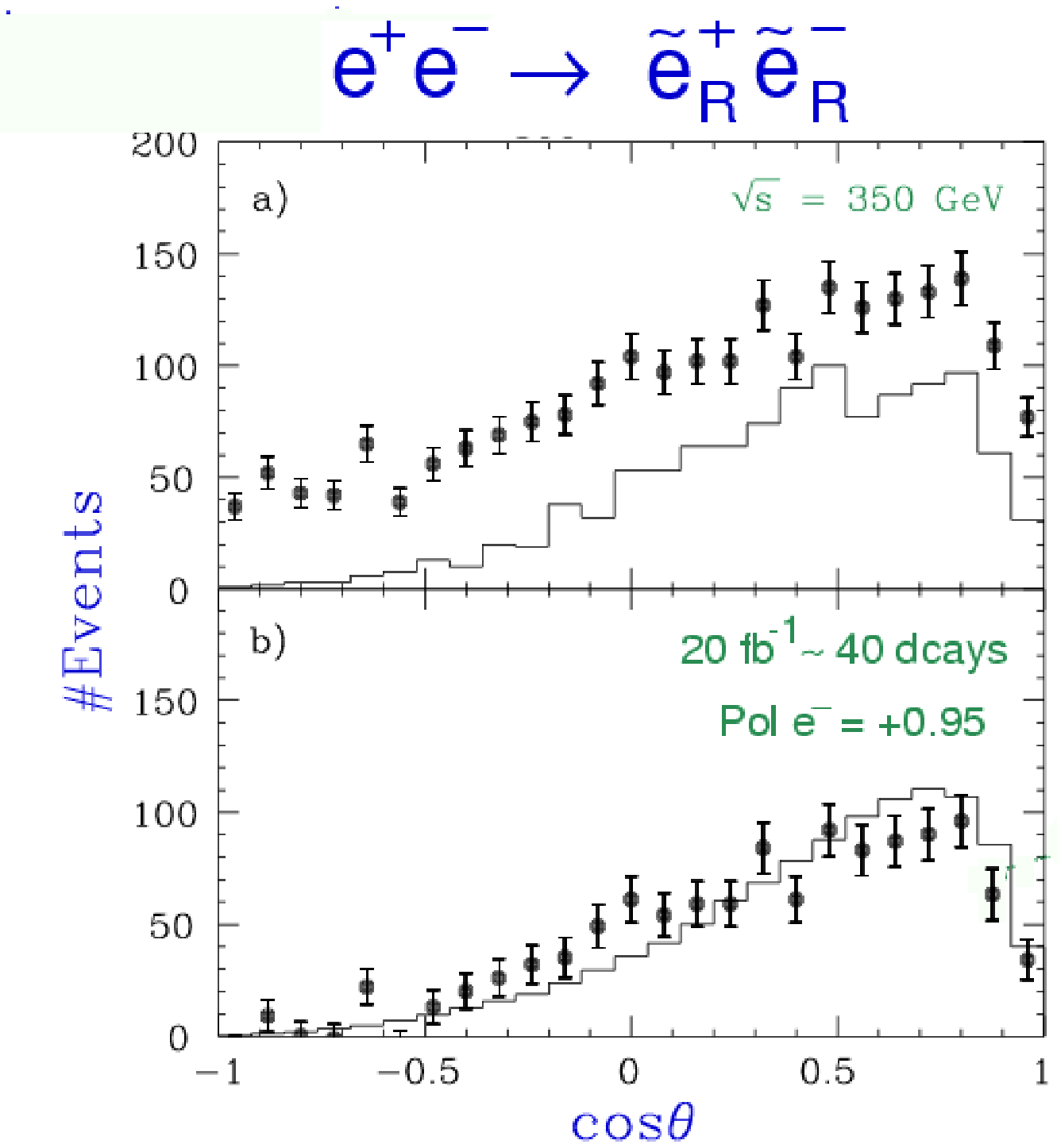}}
\end{center}
\end{figure*}

In this respect the LC is invaluable. First, it offers a complementarity with the LHC
which is better at discovering the non-coloured particles which, by the way, in many
models (unification models) have much smaller masses than the coloured-ones. Second and
most important, not only one has a far cleaner environment but one can optimize the
energy of the machine so that only very few thresholds are crossed at a time. Thus the
confusing mixing of final states with the cascade decays is avoided. There will probably
be no \susy background to \susy signals, or else one would know how to simulate the \susy
background. Third and as important is to make full use of the power of polarisation,
which takes all its meaning for a theory whose inner stucture is based on spin/chirality
symmetry! Take for instance the case of sfermions. Even in the simple case of sfermions,
\susy predicts that to each fermion chirality corresponds a sfermion. Since \susy is
broken each of these sfermion $\tilde{l}_{R,L}$ may acquire a different mass (beside a
so-called $D$-term contribution of gauge origin but involving the unknown $tg \beta$).
What is more, electroweak \sb mixes these two states, fortunately the effect is
proportional to the mass of the corresponding fermion, but involves yet two other
parameters ($\mu$ and the $A_f$ tri-linear couplings). Even in the case of the first and
second family where the latter problem is not present, one still has a few parameters to
determine. One should also make sure that one is identifying the right (correct)
sfermion. That is where polarisation comes in handy. By selecting or reconstructing the
chirality of the usual fermions one is almost directly picking up and unambiguously
identifying the appropriate sfermion because of the fact that both fermion and sfermions
share some common quantum numbers. This strategy is either not available at the LHC
(initial polarisation) or too difficult to implement (final state polarisation, as we saw
with $W$ physics). Once the identifications have been made, one can measure masses (and
possibly other parameters) and then check some mass relations without relying on any
model.\\
\begin{figure*}[htbp]
\caption{\label{smur}{\em Effect of polarization on the muon acoplanarity angle for
right-smuon and selectron production with the decay $\tilde{\mu,(e)}_R \ra \mu,(e)
\chi^0$. The slepton mass is 142GeV and the LSP is 118GeV, $\sqrt{s}=350GeV$ and ${{\cal
L}}=20fb^{-1}$. The degree of polarisation is also shown}\cite{JapanSusy}.}
\begin{center}
\vspace{-1cm}
\mbox{\epsfxsize=15cm\epsfysize=9cm\epsffile{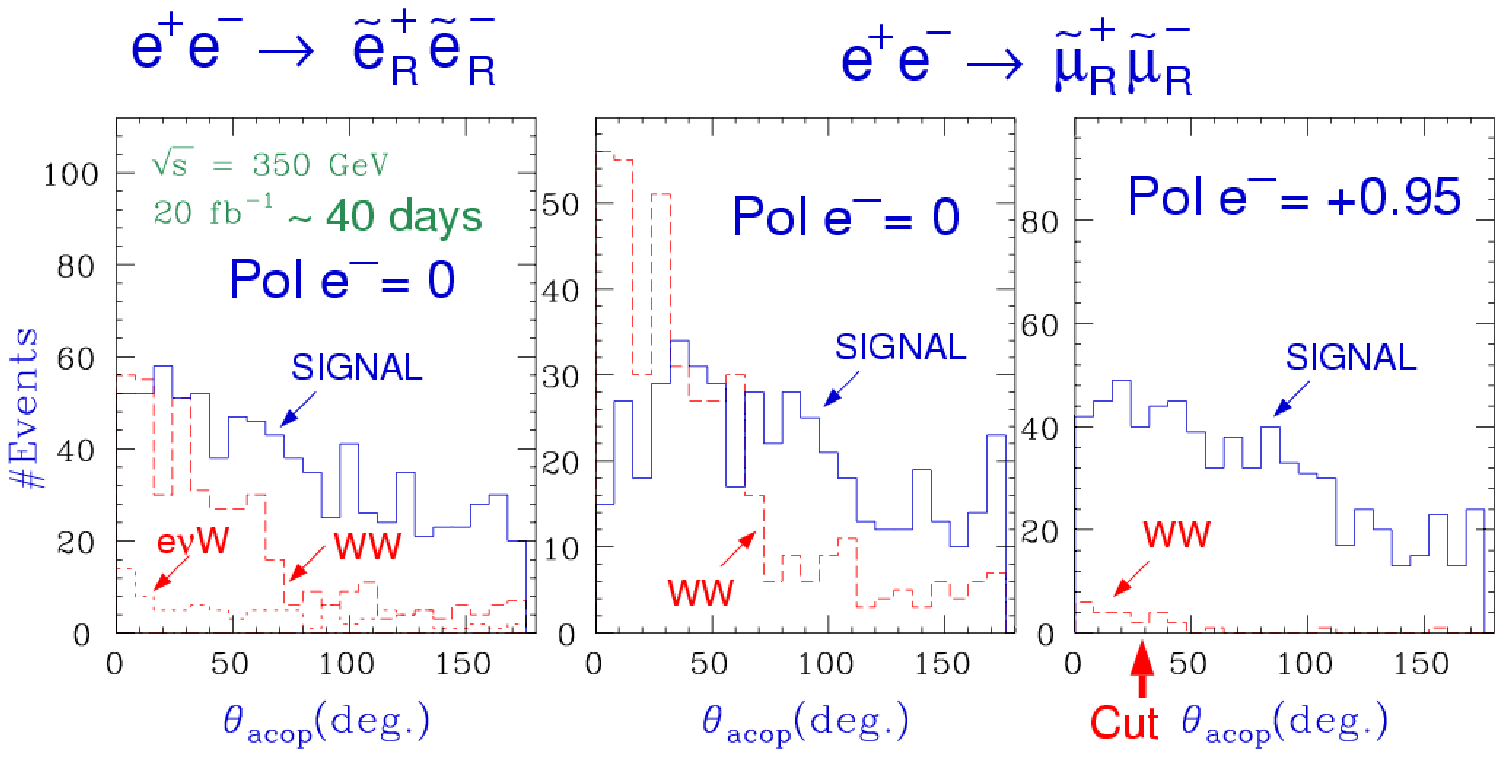}}
\end{center}
\end{figure*}
For instance, take the pair production of a right-handed smuon which most probably will
decay into the LSP neutralino and a muon. The signature is the same as that of $W$ pair
production with the $W$'s decaying into muons and neutrinos and would constitute a
formidable background. The use of polarisation becomes almost a must. First of all, $W$
pair production which is essentially an SU(2) weak process can be switched off by
choosing right-handed electrons. Indeed, at high-energy one recovers the symmetric case
where the $Z$ and $\gamma$ separate into the orthogonal $W^0$ and $B$ (hypercharge). The
former not coupling to right-handed states. On the other hand the same argument shows
that if only the hypercharge boson is exchanged and the fact that the hypercharge of the
right-hand electron is twice that of the left-handed one, right smuon production will be
four times larger than with left-handed $e^-$. Thus polarization achieves three things:
tags the nature of the smuon (right-handed) independently of how it decays, increases the
signal cross section and dramatically decreases the background. This is well rendered by
the full simulation of the Japanese group (see Fig.~\ref{smur}) which has conducted some
first-class studies\cite{JapanSusy} to which I will refer extensively. In the same figure
the case of the selectron is also shown. The latter has more background from single $W$
production that also vanish for right-handed electrons. Once the smuon production has
been optimized, one can either infer the mass from a threshold scan which is independent
of the decay or as is the case here, the measurement of the end-points of the muon energy
which give both the smuon mass and the LSP mass. A combined fit, for the case above and
for a modest luminosity $20fb^{-1}$, gives these masses at the $1\%$ level. One more
thing, to confirm the scalar nature of the smuon one can look at its angular distribution
which should show a $sin ^2\theta$ dependence. In the case of the right-handed selectron,
this will not be the case since even with a right-handed electron on has to deal with a
t-channel neutralino exchange. For the same reason as above only the bino component of
the neutralino will be selected. If this component is not negligible one should observe a
forward peak (see Fig.~\ref{selectron}). This component is a function of the gaugino
parameters $M_{1,2}$, the $\mu$ parameter and $tg \beta$. With the knowledge of
$\chi^0_1$ one can measure how much of the LSP is bino.
The mass of the right-selectron is based on the same idea as in
the smuon case. One can thus already with two processes
($\tilde{\mu}^+_R \tilde{\mu}_R^-, \tilde{e}^+_R \tilde{e}_R^- $)
test the universality of the scalar masses (at least, at this
point, for the right-sfermion masses of the first two generations)
and constrain somehow the neutralino mixing matrix. \\

As one increases the energy new thresholds may open up, for
instance the
 production of a
selectron-right with a selectron-left. This occurs only through a t-channel neutralino.
Again if one polarizes the electron to be right-handed not only one tremendously reduces
the background but also one has a better handle on the signal. In this case the
neutralino is projected onto the bino component that was present in $\tilde{e}^+_R
\tilde{e}^-_R$. Second if both selectron species decay into an electron/positron, we know
that the final electron is associated with $\tilde{e}^-_R$. The end-point energies of the
final positron will reconstruct the mass of the $\tilde{e}^+_L$. Again a precision of
$1\%$ is achieved. Note that left-handed electron polarisation allows in principle to
access the wino component of the neutralino. With these scalar masses more general mass
relations (and hence models) can be checked\cite{PeskinMorioka}.

\begin{figure*}[htbp]
\caption{\label{GUTrelation}{\em Results of a global fit using $e^+ e^-_R \ra
\tilde{e}_R^+\tilde{e}_R^-$ and $e^+ e^-_R \ra \chi^+_1 \chi^-_1$, to reconstruct
$M_1,M_2$ \cite{JapanSusy}}}
\begin{center}
\vspace{-2cm}
\mbox{\epsfxsize=9cm\epsfysize=9cm\epsffile{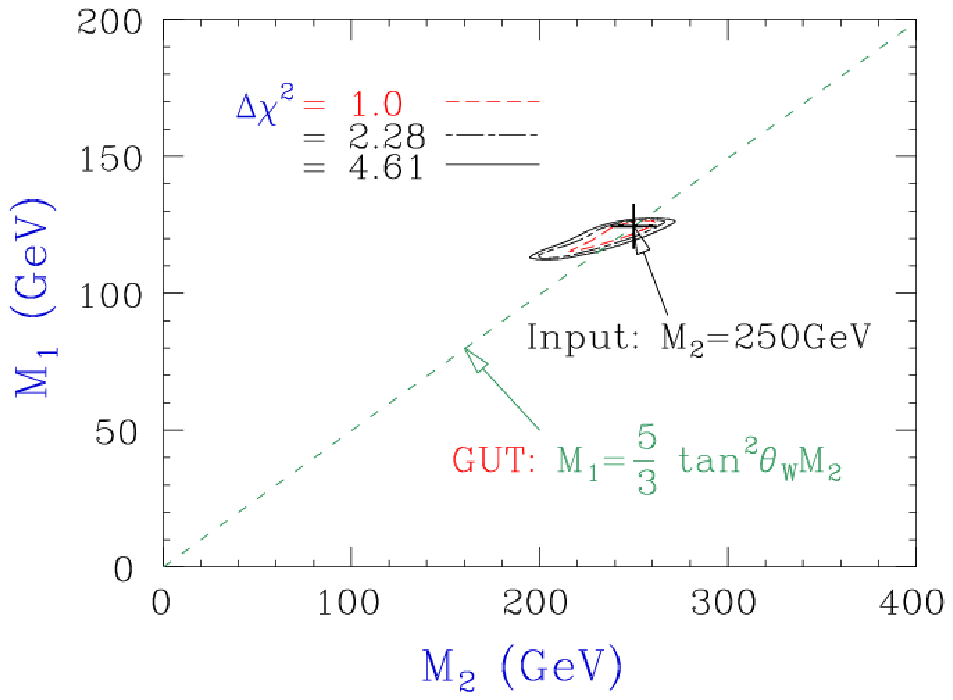}}
\vspace*{-1cm}
\end{center}
\end{figure*}

Similar analyses exploiting the power of polarisation can be done in the production of
neutralinos and charginos. Chargino pair production goes through a t-channel sneutrino
exchange as well as a s-channel $Z,\gamma$. The former can be switched off with a
right-handed electron polarization which also, through the selection of the hypercharge
component of neutral vector bosons, picks up only the higgsino component of the chargino,
therefore this polarisation alone will give us the composition of the chargino. Again
from the energy end-points of the decay products one can reconstruct the mass of the
chargino with a very good precision. By combining the information from $e^+ e^-_R \ra
\tilde{e}_R^+\tilde{e}_R^-$ and this reaction we can fit the parameters of the
chargino-neutralino mass matrix $M_1,M_2,\mu, tg\beta$ and check for the GUT relation
$M_1=\frac{5}{3} tg^2 \theta_W\; M_2$. Fig.~\ref{GUTrelation} shows the result of the fit
for a simulation based on SUGRA, but of course no SUGRA hypothesis has been made in the
fits. The results are impressive.

More can be done with the processes studied so far. With the left-handed electron
polarisation one is sensitive to the sneutrino channel and thus would measure or
constrain its mass. Needless to say that as more channels become available one can
reconstruct more fundamental parameters. Polarization will be useful if not essential.
For instance in the case of the third family one would like to measure the tri-linear
terms $A_f$ beside the $U(1)$ and $SU(2)$ scalar masses as well as $\mu$ and $-tg\beta$
if these are not already measured. Both $A_f$ and $\mu$ are contained in the angle
$\theta_f$ which mixes the left and right sfermions. This angle can be easily measured by
measuring either the cross section for a right-handed or a left-handed electron as seen
from Eq.~\ref{thetaf} ($q_f$ is the charge of the sfermion) \beqn \label{thetaf} {{\cal
M}}_L&=&{{\cal M}}_0 \biggl( |q_f| + \frac{1}{s_W^2 c_W^2}(\frac{1}{2}-s_W^2)
(\frac{1}{2} \cos^2\theta_f\;-\;|q_f| s_W^2 ) \frac{s}{s-M_Z^2} \biggr) \nonumber \\
{{\cal M}}_R&=&{{\cal M}}_0 \biggl( |q_f| - \frac{1}{c_W^2} (\frac{1}{2}
\cos^2\theta_f\;-\;|q_f| s_W^2 ) \frac{s}{s-M_Z^2} \biggr) \eeqn Decays of the third
generation sfermions will also be very informative provided one can measure
 the polarization
of the decay products, as in the case of $\tilde{\tau}$'s. These few examples make it
clear that a LC will be invaluable for precision measurements of the \susy parameters.
One last word, the \gamgamt mode will not be as helpful as the \epemt mode. The reason is
that in \gamgamt cross sections (at tree-level) are completely determined once the mass
is known or measured. Reconstruction of the parameters could only be gleaned through a
study of decays.

\section{Conclusions}
There is no doubt that the construction of a LC even if done after
the LHC will allow some crucial tests as concerns our
understanding of symmetry breaking and would very nicely
complement the LHC program. Recently the TESLA people have shown
that one can achieve even higher luminosities,
$500fb^{-1}$\cite{Tesla500fb}. This will allow even more powerful
precision tests as the ones that we went through in this summary.

{\bf Acknowlegments:} I would like to express my appreciation to the organizers for
providing us a superb atmosphere and excellent food. There could not have been a better
place than IUCCA for such a kind of Workshop. I would also like to thank the theory
division of TIFR-Mumbai for the invitation.

\small
\small

\normalsize
\end{document}